\documentclass[12pt]{iopart}

\usepackage{iopams}  
\usepackage{graphicx}

\begin{document}

\title{Determination of the Mott insulating transition
by the multi-reference density functional theory}

\author{K. Kusakabe
}

\address{
Graduate School of Engineering Science, 
Osaka University, 1-3 Machikaneyama-cho, Toyonaka, Osaka, 560-8531, Japan}  


\begin{abstract}
It is shown that a momentum-boost technique applied to 
the extended Kohn-Sham scheme enables 
the computational determination of the Mott insulating transition. 
Self-consistent solutions are given for correlated electron systems by 
the first-principles calculation 
defined by the multi-reference density functional theory, in which 
the effective short-range interaction can be determined by 
the fluctuation reference method. 
An extension of the Harriman construction is made for 
the twisted boundary condition in order to define the momentum-boost 
technique in the first-principles manner. 
For an effectively half-filled-band system, 
the momentum-boost method tells that the period of a metallic ground 
state by the LDA calculation is shortened to the least period 
of the insulating phase, indicating occurrence of 
the Mott insulating transition. 
\end{abstract}


\section{Introduction} 

Detection of the Mott insulating transition is 
a desirable function for the first-principles calculations, 
which has been demanded for years. 
A recently developed multi-reference density functional theory 
with the fluctuation reference method defines a self-consistent 
first-principles calculation, in which a short-range 
correlation effect is explicitly included.\cite{Kusakabe2,Kusakabe3,Kusakabe1} 
This technique is a generalization of the Kohn-Sham scheme of 
the electronic structure calculation.\cite{Hohenberg-Kohn,Kohn-Sham} 
Incorporation of the effective many-body system to determine 
the total energy and the single-particle charge density of 
the electronic state became possible through 
1) introduction of a fluctuation-counting term and 2) the reformulation of 
the exchange correlation energy functional as a residual exchange 
correlation energy functional. 
The effective Hamiltonian appearing in the theory is 
a kind of the Hubbard model\cite{Hubbard} or 
the Anderson model\cite{Anderson}. 
The LDA+U Hamiltonian\cite{LDA+U} is derived as 
an approximation for the residual exchange correlation 
functional.\cite{Kusakabe4} 

In this short paper, I show a technique to determine 
the Mott insulating phase in this first-principles calculation. 
The method is an application of 
the momentum boost technique known in the literature.\cite{Kohn,Kusakabe_B} 
To introduce the momentum boost method in the density functional theory, 
one needs to show the N representability. 
This is done by extending the Harriman construction\cite{Harriman,Lieb} 
for the twisted boundary condition. 
A test calculation will be shown using a simple artificial system, 
which may be represented by a one-dimensional Hubbard model. 
In the last part, I will summarize the present work. 

\section{Momentum boost technique}

I consider a Born-von-Karman boundary condition with a twist. 
Consider an array of atoms in one dimension and 
let $L$ the number of atoms in the direction, 
which is called the $x$ direction below. 
Formally using a phase $2\pi \Phi/\Phi_0$ with 
$0\le \Phi < \Phi_0$ and a unit flux $\Phi_0$, 
I introduce a twisted boundary condition in this $x$ direction. 
\begin{eqnarray}
\lefteqn{\Psi(x_1,\cdots,({\bf r}_j+La{\bf e}_x,\sigma_i),\cdots,x_N)}
\nonumber \\
&= \exp\left(2\pi i \Phi/\Phi_0 \right)
\Psi(x_1,\cdots,({\bf r}_j,\sigma_i),\cdots,x_N) , \\
&c^\dagger_{N+1,\sigma}=\exp\left(2\pi i \Phi/\Phi_0 \right)
c^\dagger_{1,\sigma} . 
\end{eqnarray}
Here, I used a combined coordinate $x_j=({\bf r}_j,\sigma_j)$ 
with the space and spin coordinates. ${\bf e}_x$ is 
a unit vector and $a$ is the lattice constant in the $x$ direction. 
$\Psi$ is a many-body wavefunction and $c^\dagger_{i,\sigma}$ 
is a creation operator defined with a properly determined 
Wannier basis $\phi_i({\bf r})$. Indeed, 
determining a Wannier transformation fixing a gauge 
in the unitary transformation,\cite{Marzari} the set of 
$\phi_i({\bf r})$ is uniquely determined in each self-consistent 
step of the extended Kohn-Sham scheme.\cite{Kusakabe3} 

Using another gauge transformation, 
the twisted boundary condition is represented 
by shifted $k$ vectors, $\displaystyle {\bf k}=(\frac{2\pi m}{La}
+\frac{2\pi\Phi}{La\Phi_0}, k_y, k_z)$, for the single-particle Bloch orbitals. 
Here $m$ is an integer within a range of $0\le m < L$. 
From the single-particle energy $\varepsilon_{n}({\bf k})$, 
one may construct a tight-binding model written in 
$c^\dagger_{i,\sigma}$ and $c_{i,\sigma}$. 

Determining the eigen energy of the effective Fermion 
system, one has an energy-flow diagram as a function of $\Phi$. 
If the system is a metal, the flow given by adiabatic connection of 
the ground state should show a long extended 
AB period, while the period has to be $\Phi_0$, when there is 
a charge gap due to formation of a Mott gap. 
Although the density-functional theory utilized in the present work 
is the ground-state formulation, one can have a signal of 
the change in the period as disappearance of a cusp in the 
lowest-energy flow. 
In case of the Mott insulator, the flux line becomes a smooth curve 
without any cusp. The appearance of the isolated lowest flow lie 
implies uniqueness of the ground state separated by a gap 
from charge excitations in the whole range of twist and 
thus indicates the formation of the Mott gap. 
If one has a difficulty in finding the change of the period 
with a three-dimensional sample, one may 
consider a one-dimensional setup using a needle like sample. 

Before discussing the result, I show that the $N$ representability of 
the present problem is guaranteed. 
In the present setup, only the boundary condition 
in the $x$ direction is a twisted one. 
Thus I consider a slice of the charge density 
$\tilde{\rho}(x) = \rho({\rm r})$ 
fixing $y$ and $z$ coordinate for simplicity. 
We are seeking for $\varphi_k(x)$ satisfying 
\begin{equation}
\tilde{\rho}(x)=\sum_k\lambda_k|\varphi(x)|^2,
\end{equation}
where $k$ is an integer, $0\le \lambda_k\le 1$ and 
$\displaystyle \sum_k \lambda_k=N$. 
In the twisted boundary condition, however, 
$\tilde{\rho}(x+La)$ obeys the periodic boundary condition as 
\begin{eqnarray}
\lefteqn{\tilde{\rho}(x+La)}\nonumber \\
&=&
\int d\sigma_1 dx_2\cdots dx_N
\Psi(({\bf r}_1+La{\bf e}_x,\sigma_1),x_2,\cdots,x_N)^* 
\nonumber \\ &&\times 
\Psi(({\bf r}_1+La{\bf e}_x,\sigma_1),x_2,\cdots,x_N)
\nonumber \\
&=&
\int d\sigma_1 dx_2\cdots dx_N
\Psi(({\bf r}_1,\sigma_1),x_2,\cdots,x_N)^*
\Psi(({\bf r}_1,\sigma_1),x_2,\cdots,x_N)
\nonumber \\
&=&
\tilde{\rho}(x).
\end{eqnarray}
Here, integration with respect to $\sigma_j$ should be interpreted as 
a summation. 
Thus, we can readily prepare the orbital 
$\varphi_k(x)$ obeying the twisted boundary condition as, 
\begin{equation}
\varphi_k(x)=\frac{\tilde{\rho}(x)}{N}^{1/2}\exp (ikf(x)+i2\pi\Phi/\Phi_0),
\end{equation}
with 
\begin{equation}
f(x)=\frac{2\pi}{N}\int_0^{La}\tilde{\rho}(x')dx'.
\end{equation}
We can immediately show that a set of $\varphi_k(x)$ forms 
complete orthonormal and that 
\begin{equation}
\sum_k\lambda_k|\varphi(x)|^2
=\frac{1}{N}\tilde{\rho}(x)\sum_k\lambda_k=\tilde{\rho}(x).
\end{equation}
Using the orbital wavefunction $\varphi_k(x)$ obeying 
the twisted boundary condition, we can show existence of 
a many-body state $|\Psi\rangle$ whose coordinate expression 
is the single Slater determinant made on $\varphi_k(x)$. 

\section{One-dimensional Hydrogen array}

As a test calculation, I consider a 
one-dimensional Hydrogen array. 
The system is denoted by an outer unit cell with ten atoms 
for the many-body calculation and an inner unit cell 
with a single atom for the single-particle problem. 
The lattice parameters for the inner cell are $a=2\AA$ and 
$b=c=10\AA$. 
The determination of $U$ may be achieved by setting 
a reference calculation.\cite{Kusakabe3} Here, to show 
the change in the period explicitly, I consider 
$U/t$ as a parameter and do only the self-consistent calculation 
for the extended-Kohn-Sham system. Here, $t$ is the value 
of the nearest-neighbor transfer parameter. 
Note that the present system is represented by a tight-binding model with 
long-range hopping terms. 
For the exchange-correlation energy functional, 
I utilized the Perdew-Zunger parameterization of 
the Ceperley-Alder diffusion Monte-Carlo data.\cite{PZ} 
The Troullier-Martins soft pseudopotential is used with 
the cut-off energy of 20 [Ry].\cite{TM} 
This setup is confirmed to be enough accurate for the discussion below 
by increasing the parameters. 
The numerical diagonalization with the Lanczos algorithm is used to 
obtain the many-body state of the first-principles Hubbard model. 

\begin{figure}
\label{Mom-boost}
\begin{center}
\includegraphics[width=7cm]{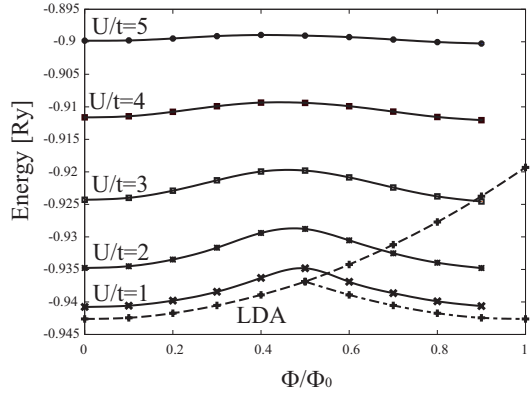}
\end{center}
\caption{
The ground state energy of a Hydrogen array system as 
a function of the phase $\Phi$ in the boundary condition. 
The system has 10 atoms. The value of $U/t$ is set to be 
$0, 1, 2, \cdots ,5$ with $t$ is the nearest-neighbor 
transfer parameter. For the case of $U=0$ (LDA calculation), 
crossing of the flow lines occur at $\Phi=\Phi_0/2$, 
while the crossing becomes anti-crossing due to the charge-gap 
formation for finite $U$. 
}
\end{figure}

The result of the momentum boost is depicted in Fig. \ref{Mom-boost}. 
For $U=0$, the LDA calculation shows a crossing in 
energy flow lines. In this case, the lowest energy flow can be 
traced, since the constrained LDA calculation fixing the 
filling of each $k$ point is available. 
Once a finite $U$ is introduced, the many-body calculation 
automatically concludes the lowest branch of the energy flows. 
In this system, we see a continuous change of the energy flow 
which has an energy gap structure at $\Phi=\Phi_0/2$. 
This result is qualitatively the same as the single-band Hubbard model 
with only the nearest neighbor hopping.\cite{Kusakabe_B} 
If the fluctuation reference method is precisely applied, 
a finite value of $U$ is expected, since the inter-atomic distance of 
$2\AA$ is in a strong-correlation regime for 
the Hydrogen molecule.\cite{Kusakabe3}
Thus the present result gives a concrete test for the method 
of determination of the Mott insulator from the first-principles. 

\section{Summary}

We have shown that the momentum boost method is 
formulated rigorously in the density functional theory. 
The N representability is shown for the twisted boundary condition. 
Using the multi-reference density functional theory (MR-DFT), 
we are able to construct an effective interacting Fermion system, 
which may undergo the Mott insulating transition. 
The momentum boost technique is applicable for this problem 
to detect the transition. 
In a realistic system, the formation of the Mott gap 
by the applied pressure or by the effective carrier doping 
would be seen as a change in the period of the lowest energy flow. 

\vspace*{5mm}

This work is partly supported by the Grand-in-Aid for 
the scientific research (No. 17064006, No. 15GS0213 and No. 19310094) and 
the 21st COE Program by the Japan Society for Promotion of Science.

\end{document}